\documentclass[aps,twocolumn,showpacs,amsmath,amssymb,floatfix,superscriptaddress]{revtex4-2}
\usepackage{amsfonts}
\usepackage{dcolumn}
\usepackage{graphicx,amssymb,amsmath}
\usepackage{bm}
\usepackage{natbib}
\usepackage{epstopdf}
\usepackage{xcolor}
\begin{document}

\title{The Statistical Mechanics of Indistinguishable Energy States and the Glass Transition }
\author{Shimul Akhanjee}
\affiliation{akhanjee.shimul@gmail.com}
\email[]{akhanjee.shimul@gmail.com}
\date{\today}

\begin{abstract}
The statistical mechanics of particles that populate indistinguishable energy substates is explored. In particular, the mathematical treatment of microstates differs from conventional statistical mechanics, where, for a given degeneracy, the energy sublevels or substates are universally treated as distinguishable and differentiated by unique quantum numbers or addressed by distinct spatial locations. Results from combinatorial counting problems are adapted to derive exact distribution functions for both classical and quantum particles at a high degeneracy limit. Quantum particles obey a nonextensive entropy $\mathcal{S} \propto \sqrt{N}$ that satisfies the area law $\mathcal{S} \propto A$ in $d=2$ bulk spatial dimensions. Classical particles exhibit a definitive glass transition similar to supercooled liquids, for which the configurational entropy vanishes below a finite temperature $T_K$.
\end{abstract}

\maketitle
\section{introduction}

Unlike a crystal, which settles into a low degeneracy periodic ground state, a glass is a low temperature $T$ phase defined by its inability to reach equilibrium on macroscopic timescales \cite{1,2,3,4,5}. The phase transition from a simple, global energy minimum to a phase defined by a complex, rugged energy landscape is one of the most important unsolved problems in physics. The broad definition of a glass includes both spin glass (SG) materials and supercooled liquids (SCLs). While SG systems exhibit a sharp phase transition to a frozen state involving quenched disorder at a specific freezing temperature $T_{SG}$, glass-forming liquids undergo an inherently nonequilibrium phase transition without explicit quenched disorder or a well-defined transition $T_g$ value \cite{1,2,6}. Both cases contain frustrated basins of attraction, or local minima that represent different amorphous structural arrangements \cite{7}. Since the particles are trapped in one of many metastable local minima, one simplified approach that will be taken here is to characterize the glass phase as the limit of highly degenerate, indistinguishable energy states. These minima are separated by high level energy barriers that grow significantly as $T$ is lowered. The glass transition is commonly understood as the point where the system can no longer hop between degenerate basins within experimental timescales, resulting in a unique type of broken ergodicity, unlike in a standard gas or liquid, where the time average of macroscopic variables equals their ensemble averages \cite{8}.

In the glassy limit, the system is confined to a single valley or a small cluster of valleys with nearly identical macroscopic properties. Distinguishing one glassy valley from another with the exact same macroscopic energy and density requires total precise knowledge of all molecular coordinates, which is effectively inaccessible. If the system cannot dynamically explore the landscape to tell the valleys apart, then the degenerate substates become operationally indistinguishable. Therefore, a reasonable question to ask is, What are the physical properties of a system where the degenerate energy substates are unlabeled to the microstate counter? Although the primary physical motivation arises from an attempt to characterize ``glassy'' phases of matter, the mathematical framework of indistinguishable energy substates is, in itself, an important foundational topic of statistical mechanics that will be explored in this paper.

\section{Phenomenological Considerations}

In glass-forming liquids, the relaxation time $\tau$ and viscosity $\eta$ are essential for understanding the nature of the transition and can be computed directly from the distribution function and the configurational entropy $\mathcal{S}_l$ \cite{1}. Both are related by $\tau = \eta / G_\infty$, where $G_\infty$ is the infinite frequency shear modulus. Typical values for $\tau$ in a liquid are $10^{-1} - 10$ s, and near $T = T_g$, $\tau$ is of the order of $10^{2} - 10^{3}$ s \cite{1}. Adam and Gibbs proposed that the relaxation of a supercooled liquid requires a cooperatively rearranging region whose size is inversely proportional to $\mathcal{S}_l$ \cite{9}. It follows that a widely used connection between $\mathcal{S}_l$ and $\tau$ is given by the Adam-Gibbs (AG) relation \cite{10},

\begin{equation}
\ln \tau \propto \frac{1}{T\mathcal{S}_l(T)}
\end{equation}
Since a glass is expected to be several orders of magnitude more viscous than liquid, it is important to establish a clear line of demarcation. Clearly, $\tau$ is a kinetic variable governed by thermally activated processes; therefore, if one considers an activation energy $E_A$, the average relaxation time of a glass should slow faster as $T$ is reduced \cite{11}. Thus, highly viscous liquids should be compared against the benchmark Arrhenius activation law \cite{10,12},
\begin{equation}
\tau \sim \tau_0 \exp\left(\frac{E_A}{k_B T}\right)
\end{equation}
Another essential concept is the Kauzmann temperature $T_K$ \cite{1,12,13}. As a liquid is supercooled, its configurational entropy $\mathcal{S}_l$ decreases drastically until it reaches the crystal threshold value. If one extrapolates this decrease, $\mathcal{S}_l$ will appear to vanish at $T = T_K$. This leads to a paradox since $\mathcal{S}_l$ would take on values lower than that of an ordered crystal. The precise resolution is still elusive but extremely important here because $\mathcal{S}_l$ will be computed explicitly for classical particles with indistinguishable energy levels and $\mathcal{S}_l$ appears to vanish precisely at some $T_K$, suggesting that the system has a glass transition. Additionally, I have determined the material dependence of $T_K$ on the single particle molecular energy bandwidth $W$ and chemical potential $\mu$.

The AG relation (1) can be used to provide an intuitive derivation of the celebrated Vogel-Fulcher-Tammann (VFT) law, which is a prevalent description of ``super-Arrhenius'' behavior near $T=T_g$ \cite{10,12,14}. In a more realistic description of a glass, the effective barrier $\Delta E$ is not constant; rather, it increases as the temperature decreases because the system must move cooperatively to relax. As the system approaches the ideal glass limit, the number of states decreases, and $\tau$ increases explosively. The VFT law is the standard empirical fit for the viscosity of fragile glass formers. Near the glass transition, it is thought that $\mathcal{S}_l(T) \approx C(1 - \frac{T_K}{T})$. Substituting this into Eq.~(1) yields the VFT form \cite{10,12}:
\begin{equation}
\tau = \tau_0 \exp\left(\frac{D T_K}{T - T_K}\right)
\end{equation}
for some material dependent parameter $D$. This expression attempts to describe the experimental observation that $\tau$ diverges because the number of available degenerate states vanishes rapidly at $T = T_K$ \cite{12}.

Before proceeding to the development of microstate counting, it is important to clarify the scope and scale of the models presented here. Since the focus of this paper is on the constraints of degenerate states, it should be emphasized that these degeneracies almost always arise as a direct consequence of a system's global symmetries. Since the exact microscopic Hamiltonian of realistic materials is influenced by quenched disorder or interactions, the true exact degeneracies might be lifted, leading to level repulsion. Therefore, the models considered here are explicitly noninteracting, clean systems that can be utilized in effective theories that capture the coarse-grained energy landscapes of the SCL phase, where the broken ergodicity confines the system to disjoint regions of phase space and the indistinguishable degenerate states represent an effective description of the system being trapped in one of these many thermodynamically equivalent degenerate basins, which is a well-established conceptual framework in the literature.

Debenedetti et al. demonstrated that the multidimensional configuration space of a SCL is tiled by basins of attraction, where the thermodynamic behavior separates into a purely configurational part and a vibrational deformation part within the basin \cite{15}. Similarly, the work by Shell et al. provides a rigorous theoretical analysis of the separation of liquid-state properties into inherent structure and vibrational components \cite{16}. This framework supports the idea that the thermodynamics of SCLs can be accurately represented by a landscape based formulation where the system's properties are derived from the distribution of these basins, directly underpinning the use of indistinguishable states.

For a more modern perspective on how these landscapes lead to ergodicity breaking in structural glasses, Scalliet et al. discussed the mechanism behind ergodicity breaking in SCL regimes, detailing how glasses evolve in an increasingly complex energy landscape featuring a large number of minima and how that relates to collective structural motion and the onset of nonergodic dynamics \cite{17}. Finally, in order to defend the idea that ergodicity breaking leads to a vastly degenerate landscape of macroscopic states, one can examine the work by He and Lubchenko in which they explicitly discuss how, in a glassy melt, the free energy surface becomes a highly degenerate landscape, leading to broken ergodicity \cite{18}. They detail that under these conditions, one tracks a large collection of distinct metastable profiles that are equilibrated regarding vibrations but not translations, serving as a coarse-grained thermodynamic description.
\section{DERIVATION OF THE DISTRIBUTION FUNCTIONS}
\subsection{Preliminaries}
Consider systems within the microcanonical ensemble having a fixed number of particles $N = \sum_j n_j$, total energy $U = \sum_j \epsilon_j n_j$ and volume $V$. The configurational entropy $\mathcal{S}_l = k_B \ln(\Omega_l)$ depends on whether the microstates arise from distinguishable particles $\Omega_{clas}(\{n_j\})$ or indistinguishable particles $\Omega_{quan}(\{n_j\})$.
\begin{equation}
\Omega_{clas}(\{n_j\}) = \prod_j \frac{t_j(n_j, g_j)}{n_j!}
\end{equation}
\begin{equation}
\Omega_{quan}(\{n_j\}) = \prod_j t_j(n_j, g_j)
\end{equation}

Evidently, the extensivity condition for distinguishable particles, $\mathcal{S}_{clas} \propto N$ is enforced by the extra $n_j!$ denominator of Eq.~(4), as prescribed by the resolution of the Gibbs paradox \cite{19,20}. The quantity $t_j(n_j, g_j)$ is explicitly a counting problem determined by the number of particles $n_j$ and degenerate energy levels $g_j$. I previously reduced the problem of determining $t_j$ to an exhaustive classification scheme of combinatorial counting problems, particularly the number of ways that one can distribute a specified number of balls into boxes, as shown in Table I, known as the 12-fold way in enumerative combinatorics \cite{21,22}.

\begin{table}[h]
\centering
\caption{The 12-fold way for $t_j$. $\left\{ \protect\begin{smallmatrix} n \\ k \protect\end{smallmatrix} \right\}$ are Stirling numbers of the second kind, $p_{\le g}(n)$ are integer partitions of $n$ into at most $g$ parts, $p_g(n)$ are integer partitions of $n$ into exactly $g$ parts. The two cases of identical $g$ with unrestricted sorting, corresponding to the identical substates for distinct $n$ (classical particles) or identical $n$ (quantum particles) are highlighted.}
\begin{tabular}{|l|c|c|c|}
\hline
$n$ and $g$ & Any sorting & Max. of 1 & Min. of 1 \\ \hline
Distinct $n$, distinct $g$ & $g^n$ & $\frac{g!}{(g-n)!}$ & $g! \left\{ \begin{matrix} n \\ g \end{matrix} \right\}$ \\ \hline
Identical $n$, distinct $g$ & $\binom{g+n-1}{n}$ & $\binom{g}{n}$ & $\binom{n-1}{g-1}$ \\ \hline
Distinct $n$, identical $g$ & $\sum_{k=1}^g \left\{ \begin{matrix} n \\ k \end{matrix} \right\}$ & $1 \text{ if } n \le g$ & $\left\{ \begin{matrix} n \\ g \end{matrix} \right\}$ \\ \hline
Identical $n$, identical $g$ & $p_{\le g}(n)$ & $1 \text{ if } n \le g$ & $p_g(n)$ \\ \hline
\end{tabular}
\end{table}

It follows that the next section will involve the application of this mathematical framework to quantum and classical particles with indistinguishable energy sublevels. It should be made clear that the particles can distinguish between the single particle $\epsilon_j$ levels but not among the $g_j$-fold degenerate sublevels. I do not explicitly include the effects of either disorder or interactions. Combinatorically, this corresponds to mapping $n$ distinct or identical balls into $g$ identical boxes. Since I am interested in unrestricted sorting for indistinguishable energy sublevels, the first column and the last two rows of Table I will be explored below as separate cases.

\subsection{Classical particles with indistinguishable energy sublevels}
I start with
\begin{equation}
t_j(n_j, g_j) = \sum_{k=1}^{g_j} \left\{ \begin{matrix} n_j \\ k \end{matrix} \right\}
\end{equation}
where, in combinatorics, the brackets $\left\{ \begin{matrix} n_j \\ k \end{matrix} \right\}$ represent the Stirling numbers of the second kind \cite{23}. Equation (6) describes the number of ways to partition $n_j$ distinct particles into at most $g_j$ indistinguishable substates. Since the states can be empty without any restrictions on the occupancy, the sum over the possible number of occupied substates is taken to a maximum value of $g_j$. To find the distribution function $n(\epsilon)$, Eq.~(4) is applied to Eq.~(6), and $\mathcal{S}_l$ is maximized, with the usual Lagrange multipliers $\alpha$ and $\beta$ to enforce the conservation of $N$ and $U$ \cite{20}. Thus, I evaluate $n(\epsilon)$ in two distinct thermodynamic regimes.

\subsubsection{Small degeneracy $(n_j \gg g_j)$}
In this regime, the number of particles vastly outnumbers the available substates. Furthermore, since the substates are indistinguishable, the particles are forced to occupy all available states, and the sum is heavily dominated by the maximum number of partitions, $k = g_j$. The asymptotic behavior of Eq.~(6) when $n_j \gg g_j$ is
\begin{equation}
t_j(n_j, g_j) \approx \frac{g_j^{n_j}}{g_j!}
\end{equation}
The configurational weight for level $j$ after using Stirling's approximation $\ln x! \approx x \ln x - x$ becomes
\begin{equation}
\ln\left(\frac{t_j}{n_j!}\right) \approx \ln\left(\frac{g_j^{n_j}}{g_j! n_j!}\right) \approx n_j \ln g_j - \ln g_j! - (n_j \ln n_j - n_j).
\end{equation}
After one takes the derivative with respect to $n_j$ to maximize $\mathcal{S}_l$,
\begin{equation}
\frac{\partial \mathcal{S}}{\partial n_j} = k_B (\ln g_j - \ln n_j) = -k_B \ln\left(\frac{n_j}{g_j}\right)
\end{equation}
and equates the expression above to the Lagrange multipliers $\alpha + \beta\epsilon_j$,
\begin{equation}
-k_B \ln\left(\frac{n_j}{g_j}\right) = \alpha + \beta\epsilon_j.
\end{equation}
Subsequently, one can make the usual substitutions: $\alpha \equiv -\beta\mu$, $\beta \equiv 1/(k_B T)$, and the fugacity $z \equiv e^{\mu/(k_B T)}$. This leads to the expression
\begin{equation}
n_j(\epsilon_j) = g_j z e^{-\frac{\epsilon_j}{k_B T}}.
\end{equation}
As expected, low degeneracies correspond to the $\epsilon_j \ll k_B T$ limit, and one exactly recovers the standard Maxwell-Boltzmann statistics. This result has the simple interpretation that heavily populated systems have a low probability of having empty states. Thus, the effect of the indistinguishability of the $g_j$ states will merely reduce the total phase space volume by an overall global permutation factor $1/g_j!$. This shifts the absolute zero of $\mathcal{S}$ but does not alter the relative shape of the distribution function.

\subsubsection{Large degeneracy $(g_j \ge n_j)$}
The physical implications of the large degeneracy regime are striking, and a detailed discussion of its importance to the glass problem will be provided in the next section. When there are more indistinguishable states than particles, the upper limit of $g_j$ becomes irrelevant, and the sum over Stirling numbers yields the Bell number $B_n$, which counts the total number of partitions of a set of $n$ distinct elements,
\begin{equation}
t_j = B_{n_j}
\end{equation}
After the known asymptotic expansion for $B_n$ for large $n$ is applied,
\begin{equation}
\ln B_n \approx n \ln n - n \ln(\ln n) - n.
\end{equation}
The configurational weight becomes
\begin{align}
\ln\left(\frac{B_{n_j}}{n_j!}\right) &\approx [n_j \ln n_j - n_j \ln(\ln n_j) - n_j] - (n_j \ln n_j - n_j) \nonumber \\
&= -n_j \ln(\ln n_j).
\end{align}
If one takes the derivative with respect to $n_j$,
\begin{equation}
\frac{\partial \mathcal{S}}{\partial n_j} \approx k_B\left(-\ln(\ln n_j) - \frac{1}{\ln n_j}\right) \approx -k_B \ln(\ln n_j),
\end{equation}
and assumes the term $1/\ln n_j$ vanishes for large $n_j$, the same steps as before can be followed:
\begin{equation}
-\ln(\ln n_j) = \frac{\epsilon_j - \mu}{k_B T}
\end{equation}
After one exponentiates twice, the final expression for the distribution function becomes
\begin{equation}
n_j(\epsilon_j) = \exp\left(z e^{-\frac{\epsilon_j}{k_B T}}\right).
\end{equation}

Equation (17), while keeping $z > 0$, does not have the exact form of any previously known equation in the literature. Other double-exponential distributions such as the Gompertz and Gumbel functions have different acceptable ranges of parameters \cite{24,25}. In particular, the Gumbel cumulative distribution function (CDF), is known as the type I extreme value distribution, which is used to model the maximum values of random variables \cite{25}. It takes on the general mathematical form
\begin{equation}
F(x; \mu, \beta) = \exp\left(-e^{-(x - \mu)/\beta}\right).
\end{equation}
The clear difference is that the Gumbel CDF features a double negative exponential, corresponding to the condition $z = -1$, which is physically impossible in most systems. A Gumbel CDF must approach 1 as $x \to \infty$ and 0 as $x \to -\infty$. Moreover, in Eq.~(17) $n_j$ actually grows as $\epsilon_j \to 0$ or if $z > 0$, which is more commonly observed with occupation numbers rather than a normalized probability measure.

On the other hand, the Gompertz distribution is typically used to model mortality rates or growth \cite{24}. Its probability density function has the form
\begin{equation}
f(x; \eta, b) = b \eta e^{bx} \exp\left[-\eta(e^{bx} - 1)\right].
\end{equation}
It appears to have a structure similar to the Gumbel CDF, as the Gompertz requires a negative sign in the exponent of the outer exponential to ensure that the probability decays to zero. Clearly, Eq.~(17) is an asymptotic result, describing a system where the states are occupied in a way that diverges or saturates differently than in a survival model. Moreover, its double-exponential structure produces a highly exotic thermodynamic environment that deviates wildly from classical Maxwell-Boltzmann or standard quantum distributions.

An obvious point of distress concerning the $g_j \ge n_j$ asymptotic limit is at high energies, where there is a saturation of states $n_j \to 1$. Physically, this implies a permanent, pressurized background occupation, where even the highest, most inaccessible energy states will be populated by exactly one particle on average. The total solution of Eq.~(6) would need to be studied numerically to fully understand this $g_j$ driven crossover since the $\epsilon_j \ll k_B T$ limit described by Eq.~(11) is well behaved at high energies. Since the actual number of indistinguishable substates $g_j$ naturally decreases at high energies, which is common for bound states or specific molecular potentials, the system will naturally exit the large degeneracy regime at the highest energies. The full theory would demonstrate a crossover back to a low-occupation, Maxwell-Boltzmann regime before hitting the saturation point, and a numerical interpolation between the two regimes would likely show $n_j$ smoothly decaying to zero. Provisionally, realistic materials can be approximated by a strict, finite bandwidth on their single particle energy states, eliminating the need for an infinite reservoir of particles.

\subsection{Quantum particles with indistinguishable energy sublevels}
If both the particles and the substates are indistinguishable, then the distribution of $n_j$ identical particles into $g_j$ identical substates with no occupancy restriction is equivalent to integer partitioning. The quantity of interest is
\begin{equation}
t_j = p_{\le g_j}(n_j)
\end{equation}
which is the number of integer partitions of $n_j$ into at most $g_j$ parts. In order to obtain analytical results, we must evaluate $\mathcal{S}_j = k_B \ln p_{\le g_j}(n_j)$ for large $n_j$, which branches into two asymptotic regimes which depend on the limits taken on degeneracy $g_j$ like before for the classical particles.

\subsubsection{Small degeneracy $(g_j \ll n_j)$}
If the degeneracy $g_j$ is much smaller than $n_j$, then the asymptotic expansion for restricted partitions behaves as
\begin{equation}
p_{\le g_j}(n_j) \approx \frac{n_j^{g_j - 1}}{g_j!(g_j - 1)!}.
\end{equation}
which simplifies to $\ln t_j \approx (g_j - 1) \ln n_j$. If one solves for $n_j$, the distribution function resembles the classical $\epsilon_j \ll k_B T$ limit taken on quantum systems, which is similar in form to the Rayleigh-Jeans law,
\begin{equation}
n_j(\epsilon) = \frac{k_B T(g_j - 1)}{\epsilon_j - \mu}.
\end{equation}

\subsubsection{Large degeneracy $(g_j \ge n_j)$}
If the number of substates is large, the restriction $g_j$ becomes irrelevant, and $p_{\le g_j}(n_j) \approx p(n_j)$, which is also known as the unrestricted partition function. Therefore, one can make use of the Hardy-Ramanujan asymptotic formula \cite{26,27},
\begin{equation}
\ln t_j \approx \pi \sqrt{\frac{2n_j}{3}},
\end{equation}
and can carry out entropy maximization,
\begin{equation}
\frac{\partial \mathcal{S}}{\partial n_j} = k_B \frac{\partial}{\partial n_j}\left(\pi\sqrt{\frac{2n_j}{3}}\right) = \frac{k_B \pi}{\sqrt{6n_j}}.
\end{equation}
If one equates this to the Lagrange multipliers and solves for $n_j$, the distribution function is
\begin{equation}
n_j(\epsilon) = \frac{\pi^2(k_B T)^2}{6(\epsilon_j - \mu)^2},
\end{equation}
having the form of an inverse squared distribution, reflecting the massive combinatorial growth of partitions for large numbers. Quantum particles occupying indistinguishable energy sublevels are a highly exotic statistical system where the phase space volume is drastically compressed compared to standard quantum statistical mechanics. Furthermore, in standard thermodynamics, the entropy is extensive, $\mathcal{S} \propto N$. However, in the high-degeneracy regime governed by the Hardy-Ramanujan formula, the entropy scales as
\begin{equation}
\frac{\mathcal{S}}{k_B} = 2\pi \sqrt{\frac{\kappa N}{6}}
\end{equation}
where
\begin{equation}
\kappa = \frac{\left( \int_{\Delta}^{\Lambda} \frac{\rho(\epsilon)}{(\epsilon - \mu)} d\epsilon \right)^2}{\int_{\Delta}^{\Lambda} \frac{\rho(\epsilon)}{(\epsilon - \mu)^2} d\epsilon}
\end{equation}
This implies that the additivity of macroscopic subsystems breaks down entirely. Since Eq.~(25) is an inverse squared power law, the particles have a much higher probability of occupying excited states. Apparently, there are severe divergences when macroscopic thermodynamic quantities are calculated without proper cutoffs. By assuming $\mu = 0$ and introducing an allowed energy band $\epsilon \in [\Delta, \Lambda]$, one can analyze the stability of the system under different dispersion relations. Starting with a system with a constant density of states $\rho(\epsilon) = \rho_0$,
\begin{equation}
N = \frac{\pi^2 k_B^2 T^2 \rho_0}{6} \left(\frac{1}{\Delta} - \frac{1}{\Lambda}\right),
\end{equation}
\begin{equation}
U = \frac{\pi^2 k_B^2 T^2 \rho_0}{6} \ln\left(\frac{\Lambda}{\Delta}\right).
\end{equation}
Accordingly, $U$ contains a logarithmic ultraviolet divergence, with a specific heat that is linear with temperature, $C_V \propto T$.

In $d=3$ spatial dimensions, the distinction between massive (nonrelativistic) and massless (relativistic) particles will affect the infrared stability of the system. For massive particles, $\rho_{NR}(\epsilon) \propto \sqrt{\epsilon}$, causing the $N$ integrand to scale as $\epsilon^{-3/2}$, leading to a divergence as $\Delta \to 0$, which suggests that the system is unstable against low energy particle clusters. Although the standard boson gas exhibits Bose-Einstein condensation where the particles collapse into the ground state energy, the unrestricted, indistinguishable substate model allows for a critical occupation away from the ground state. The energy substates themselves offer no distinct label to differentiate arrangements, and therefore, the system minimizes its free energy by exhibiting extreme macroscopic bunching, where fluctuations in particle number per energy level $\Delta n^2$ are massive, implying that the system behaves more like a single, macroscopic, collective excitation with a finite kinetic energy.

Conversely, for $d=3$ massless, relativistic particles (e.g., photons, Weyl fermions), $\rho_{Rel}(\epsilon) = B\epsilon^2$, where $B = V/[2\pi^2(\hbar c)^3]$, the $1/\epsilon^2$ dependence of Eq.~(25) is suppressed.
\begin{equation}
N = \frac{\pi^2 k_B^2 T^2 B}{6} (\Lambda - \Delta),
\end{equation}
\begin{equation}
U = \frac{\pi^2 k_B^2 T^2 B}{12} (\Lambda^2 - \Delta^2).
\end{equation}
Therefore, the relativistic system is intrinsically infrared stable; that is, in the limit $\Delta \to 0$ $N$ is well behaved, although the ultraviolet divergence in $U$ remains, necessitating a high energy cutoff such as a Planck scale limit. Since the relationship $N \propto T^2$ always holds, it can be concluded that the scaling form $\mathcal{S} \propto \sqrt{N}$ holds regardless of the dispersion relation.

Next, it is important to highlight a thermodynamic connection that quantum particles with indistinguishable substates have with black holes, namely, the celebrated area law, which states that $\mathcal{S} \propto A \propto L^{d-1}$ \cite{28,29}. In the quantum system considered here $N \propto L^d$, which implies that the combinatorial entropy scales geometrically as $\mathcal{S} \propto \sqrt{N} \Rightarrow \mathcal{S} \propto L^{d/2}$. If one equates the bulk combinatorial entropy to the boundary area law ($L^{d/2} = L^{d-1}$), it is satisfied at dimension $d=2$. These features may be combined with the infinite symmetries of a two-dimensional conformal field theory (CFT)$_2$, which are governed by the Virasoro algebra, whose generators $L_{-n}$ can possibly act as creation operators for indistinguishable harmonic oscillator modes \cite{30}. Calculating the microstates of the CFT at high energy levels reduces identically to the combinatorial integer partition problem that I have studied here \cite{31}. Conveniently, the high energy asymptotic for the density of states of the CFT$_2$ agrees with Eq.~(26), as it is structurally identical to the Cardy formula:
\begin{equation}
\mathcal{S} = 2\pi\sqrt{\frac{c N}{6}}
\end{equation}
for a central charge $c = \kappa$ that depends on the form of $\rho(\epsilon)$ and the cutoff parameters \cite{32}.

\section{THE GLASS TRANSITION}
\subsection{The entropy $\mathcal{S}$ and heat capacity $C_\mu$}
Leaving aside the exotic quantum system, the remaining discussion will focus on the classical system at high degeneracy governed by Eq.~(17) and how it relates to the phenomenology of the SCL glass phases. Therefore, it is necessary to study the precise thermodynamic behavior of the total configurational entropy $\mathcal{S}$ and the heat capacity $C_\mu$. I start from the grand potential $\Phi$, which is obtained by integrating the expression $n_j = -\frac{\partial\Phi_j}{\partial\mu}$ with Eq.~(17),
\begin{align}
\Phi &= -\sum_j \int n_j(\mu) d\mu \nonumber \\
&= -\sum_j \int \exp\left(e^{(\mu - \epsilon_j)/(k_B T)}\right) d\mu \nonumber \\
&= -k_B T \sum_j \text{Ei}\left(e^{(\mu - \epsilon_j)/(k_B T)}\right)
\end{align}
where $\text{Ei}(x)$ is the exponential integral \cite{23}. Next, the configurational entropy is given by
\begin{align}
\mathcal{S} &= -\left(\frac{\partial\Phi}{\partial T}\right)_{V,\mu} \nonumber \\
&= k_B \sum_j \text{Ei}\left(e^{\frac{\mu - \epsilon_j}{k_B T}}\right) - \frac{1}{T} \sum_j (\mu - \epsilon_j)\exp\left(e^{\frac{\mu - \epsilon_j}{k_B T}}\right) \nonumber \\
&= k_B \sum_j \left[\text{Ei}(y_j) - e^{y_j} \ln y_j\right],
\end{align}
where the scaling parameter is defined as $y_j = \exp\left(\frac{\mu - \epsilon_j}{k_B T}\right)$. Subsequently, the heat capacity at constant $\mu$ is
\begin{equation}
C_\mu = T\frac{\partial \mathcal{S}}{\partial T} = \frac{1}{k_B T^2} \sum_j (\epsilon_j - \mu)^2 (\Delta n_j)^2,
\end{equation}
with the variance in the energy state occupations defined as
\begin{equation}
(\Delta n_j)^2 = k_B T \frac{\partial n_j}{\partial\mu} = y_j \exp(y_j).
\end{equation}

\subsection{The Kauzmann temperature $T_K$}
\subsubsection{Asymptotic limit $T \to 0$}
Now, I will examine the asymptotic limit $T \to 0$ and attempt to solve for $T_K$. For discrete energy levels, it is convenient to separate the ground state $j=0$ from the excited states $j > 0$ and define an energy gap $\Delta\epsilon_j = \epsilon_j - \epsilon_0 > 0$. As $\mu \to \epsilon_0$, the ground state $N_0 = e^{y_0}$ is occupied by more particles. A more suitable form of the scaling parameter becomes
\begin{equation}
y_j = \exp\left(\frac{\mu - \epsilon_0}{k_B T} - \frac{\Delta\epsilon_j}{k_B T}\right) = y_0 e^{-\Delta\epsilon_j / k_B T}.
\end{equation}
In order to extract the critical behavior of $\mathcal{S}$, the primary task is to analyze the function $f(y) = \text{Ei}(y) - e^y \ln y$ and study the behavior of both the ground state and excited states when $y \to 0$ and $T \to 0$. In this limit, one can apply the series expansion $\text{Ei}(y) = \gamma + \ln y + y + \mathcal{O}(y^2)$ and $e^y \approx 1 + y$,
\begin{equation}
f(y_j) \approx \gamma + y_j - y_j \ln y_j.
\end{equation}
where $\gamma \approx 0.577$ is the Euler-Mascheroni constant \cite{27}, and after Eq.~(37) is substituted into (38), the leading contribution to $\mathcal{S}_j$ as $T \to 0$ is dominated by the energy gap,
\begin{equation}
\mathcal{S}_j \approx k_B\left[\gamma + y_0 \left(\frac{\Delta\epsilon_j}{k_B T}\right) e^{-\Delta\epsilon_j / k_B T}\right].
\end{equation}
For $j=0$ one can apply the asymptotic expansion $\text{Ei}(y) \approx \frac{e^y}{y}$
\begin{align}
\mathcal{S}_0 &= k_B \left[\text{Ei}(\ln N_0) - N_0 \ln(\ln N_0)\right] \nonumber \\
&\approx k_B \left[\frac{N_0}{\ln N_0} - N_0 \ln(\ln N_0)\right].
\end{align}
Apparently, since $N_0 \ln(\ln N_0)$ grows much faster than $N_0 / \ln N_0$, $\mathcal{S}_0$ plunges into a macroscopically large negative value at some unphysical threshold value of $T$. Therefore, to find a meaningful, finite bound for $T_K$, I will abandon the $T \to 0$ discrete ground state approximation. Instead, I will consider a continuous density of states $\rho(\epsilon)$ that will allow the particles to distribute over a macroscopic energy band, preventing the mathematical singularity of a single state condensation.

\subsubsection{Finite $T_K$}
In the continuum limit the total number of states is $\mathcal{N} = \int \rho(\epsilon) d\epsilon$, and therefore, the discretized entropy of Eq.~(34) is replaced by the following continuous entropy:
\begin{equation}
\mathcal{S} = k_B \int \rho(\epsilon) \left[\text{Ei}(y(\epsilon)) - e^{y(\epsilon)}\left(\frac{\mu - \epsilon}{k_B T}\right)\right] d\epsilon.
\end{equation}
To evaluate this near a finite $T_K$, we exploit the pseudo-Fermi surface of the high energy occupied states at $\epsilon = \mu$ and $y = 1$. Consequently, Eq.~(41) can be split into two distinct thermodynamic regimes that compete with one another; namely, the transition occurs when the $\text{Ei}(y)$ term is completely eclipsed by the negative $e^y \ln y$ term. The high energy contributions, $\epsilon > \mu$ and $y(\epsilon) \to 0$ are given by the expression
\begin{equation}
\mathcal{S}_{high} \approx k_B \int_{\mu}^{\infty} \rho(\epsilon) \gamma d\epsilon \approx k_B \gamma \mathcal{N}_{>\mu}.
\end{equation}
The saturation of $n_j \to 1$ acts like a surjective constraint, forcing a minimum occupancy across the entire spectrum and acting as a reservoir of positive structural entropy. On the other hand, the low energy states, $\epsilon < \mu$ and $y(\epsilon) \gg 1$ yield the negative entropy contribution,
\begin{align}
\mathcal{S}_{low} &\approx -k_B \int_{-\infty}^{\mu} \rho(\epsilon) e^{y(\epsilon)}\left(\frac{\mu - \epsilon}{k_B T}\right) d\epsilon \nonumber \\
&\approx -k_B N \left(\frac{\mu - \langle \epsilon \rangle}{k_B T}\right).
\end{align}
where $n(\epsilon) = \rho(\epsilon)e^{y(\epsilon)}$ leads to an average condensation energy $\langle \epsilon \rangle$, with the assumption that at lower $T$ the vast majority of particles are condensed with energies below $\mu$. The final equation, $\mathcal{S}_{high} + \mathcal{S}_{low} = 0$ is left to be solved:
\begin{equation}
k_B \gamma \mathcal{N}_{>\mu} - k_B N \left(\frac{\mu - \langle \epsilon \rangle}{k_B T_K}\right) \approx 0,
\end{equation}
yielding a closed form expression for the Kauzmann temperature,
\begin{equation}
T_K \approx \frac{N}{\mathcal{N}_{>\mu}} \frac{(\mu - \langle \epsilon \rangle)}{k_B \gamma}.
\end{equation}

Equation (45) is remarkably simple, yet profound, revealing much needed insight into the microscopic origins of the glass transition. $T_K$ scales with $N / \mathcal{N}_{>\mu}$, which is the ratio of the total number of particles to the number of states with energies $\epsilon > \mu$. The denser the system is packed relative to the available high-energy states, the higher $T_K$ is. Additionally, $T_K$ scales with the condensate depth $|\langle \epsilon \rangle - \mu|$, which causes $T_K$ to increase when most of the particles are trapped in deep energy wells far below $\mu$. Coincidently, this entropy crisis is mathematically similar to one displayed by Derrida's random energy model (REM), which is an SG system with $\mathcal{S}_R(E) = k_B(N \ln 2 - \frac{E^2}{NJ^2}) = 0$ \cite{33}. In the REM the critical distribution of the ground state energy transitions from a Gaussian, self-averaging distribution to a phase entirely dominated by the Gumbel distribution in Eq.~(18), having a form similar to Eq.~(17), as discussed earlier.

It is important to note that the $\mu$ value of atoms in a liquid is typically measured indirectly by determining the equilibrium vapor pressure $P_i$ of the component above the liquid, defined as the partial molar Gibbs energy $\mu_i = (\partial G/\partial n_i)_{T,P}$ \cite{12,34}. However, $\mu$ measurements in nonequilibrium and amorphous glass phases remain challenging. Instead, researchers use estimates that derive from advanced numerical methods such as molecular dynamics simulations and the Jarzynski relation to compute thermodynamic potentials and $\mu$ in the glassy regime \cite{35}. Furthermore, measuring the atomic single particle energy spectrum $\langle \epsilon \rangle$ and kinetic energy of liquid and glass phases can be achieved with deep inelastic neutron scattering \cite{36}. This technique allows the momentum distribution and the average kinetic energy of individual atoms to be determined through the glass phase transition. Since $C_\mu$ scales with $y \exp(y)$, it explodes superexponentially below $T_K$. The crossover between indistinguishable state statistics and Maxwell-Boltzmann statistics produces a pronounced nonlinearity in $C_\mu$, resembling the thermodynamic anomalies observed in supercooled liquids near the glass transition.

\subsubsection{$\rho(\epsilon) = \rho_0$ flat-band approximation}
One can go further in seeking a simple, closed form expression for $T_K$ and apply a constant density of states $\rho(\epsilon) = \rho_0$, also known as the flat-band approximation, to the analysis in the preceding section \cite{37}. To ensure the $n_j \to 1$ tail does not produce an infinite baseline of particles, the states are restricted to a finite bandwidth $\epsilon \in [0, W]$. This implies that the total number of available states is $\mathcal{N} = \rho_0 W$. Integrating the $\epsilon > \mu$ states will act as a background contribution to the positive entropy reservoir:
\begin{equation}
\mathcal{S}_{high} = k_B \int_{\mu}^{W} \rho_0 \gamma d\epsilon = k_B \gamma \rho_0 (W - \mu).
\end{equation}
It follows that the $\epsilon < \mu$ states contribute to the negative part of the configurational entropy,
\begin{align}
\mathcal{S}_{low} &= -k_B \int_{0}^{\mu} \rho_0 \left(\frac{\mu - \epsilon}{k_B T}\right) \exp\left(e^{\frac{\mu - \epsilon}{k_B T}}\right) d\epsilon \nonumber \\
&= -k_B \rho_0 k_B T \int_{0}^{\mu/k_B T} x \exp(e^x) dx.
\end{align}
In order to evaluate this expression near $T_K$, the extreme double-exponential weighting ensures that the area spanned by the integral is almost entirely dominated by the upper boundary $X = \frac{\mu}{k_B T}$ near the ground state $\epsilon \to 0$. If one substitutes $u = e^x$ and integrates by parts, the leading term at the boundary is exactly $X e^{-X} e^{e^X}$,
\begin{equation}
\mathcal{S}_{low} \approx -k_B \rho_0 \mu e^{-\mu/k_B T} \exp\left(e^{\mu/k_B T}\right).
\end{equation}
As before, the equation $\mathcal{S}_{high} + \mathcal{S}_{low} = 0$ becomes
\begin{equation}
\ln[\gamma(W - \mu)] = \ln \mu - \frac{\mu}{k_B T_K} + e^{\mu/k_B T_K}.
\end{equation}
Since the system is heavily condensed at $T_K$, the $e^{\mu/k_B T_K}$ term is massive and dominates the right side of Eq.~(49). After making this approximation and solving for $T_K$ we have the expression
\begin{equation}
T_K \approx \frac{\mu}{k_B \ln\left[\ln\left(\frac{\gamma(W - \mu)}{\mu}\right)\right]}.
\end{equation}
This approximation demonstrates some important qualitative features of $T_K$ that should survive in the presence of additional band curvature effects, such as the dependence of $T_K$ on $W$, which is intrinsically related to the density of higher energy states within the liquid. In order to verify this relationship in real materials, compressing the liquid or any thermodynamic process that broadens $W$ could increase the density of $\epsilon > \mu$ states and lower $T_K$. A higher $T_K$ arises only when the system is heavily condensed relative to a scarce number of $\epsilon > \mu$ states. Since the model exhibits strong dependence on $\mu$, which controls the degree of clustering in the $g \gg n$ regime, increasing $\mu$ enhances the low energy occupation and produces sharper $C_\mu$ anomalies, suggesting a direct connection between the microscopic degeneracy structure and macroscopic fragility.

\section{CONCLUSION}
In summary, the basic question of how the distinguishability of degenerate energy substates affects the macroscopic thermodynamics of both classical and quantum systems was addressed by a mathematically rigorous approach involving enumerative combinatorics. The primary attribute of these systems is the erasure of the identity of the substate within a degenerate level, and therefore, the only physically meaningful information arises from the partitioning of the particles among the primary energy levels $\epsilon_j$. The resulting physics was not developed as a simple variation of existing statistical mechanical theories, and unlike previous attempts to expand the traditional realm of distribution functions, the results presented here were not derived as a special case of parastatistics, anyons in higher dimensions, or $q$ deformations; rather, they were generated from a fundamentally new set of combinatorial constraints \cite{38,39,40}. For both quantum and classical cases, the crossover and possible sharp transition of the ground state as a function of the degeneracy $g$ need to be studied farther away from the asymptotic range.

The nonextensive $\mathcal{S}$ scaling for indistinguishable quantum states should not be perceived as a mere statistical anomaly. The natural infrared stability of the relativistic dispersion, combined with the geometric scaling in a bulk, $d=2$ spatial system, will be of interest for research concerning the Virasoro symmetry structure of CFT$_2$. On the other hand, for classical systems, a double exponential distribution and the hyper-Arrhenius vanishing of $\mathcal{S}_l$, known as the Kauzmann crisis, were derived from the microcanonical ensemble. This robust signature of an SCL thermal profile perfectly mimics the molecular structural arrest of an SCL crossing its $T_g$. As a consequence of the particles being distinguishable while the degenerate substates are not, the system minimizes its free energy by lumping particles into massive clusters given that one cluster of particles in an unlabeled energy sublevel counts as just one state, avoiding the entropic penalty of distributing them across the other substates. Physically, this extreme reduction in microstate multiplicity provides a direct, combinatorial mechanism for $\mathcal{S}_l$ to collapse and vanish at $T_K$. Last, the 12-fold way is the combinatorial foundation for the microstate classification, a mathematical structure that compliments the limitations of existing theoretical methods used to study the glass transition.

\section*{DATA AVAILABILITY}
There are no publicly available research data or software supporting this manuscript. Requests for further information or data should be sent to the authors.


\begin{thebibliography}{99}
\bibitem{1} J. C. Dyre, Colloquium: The glass transition and elastic models of glass-forming liquids, Rev. Mod. Phys. 78, 953 (2006).
\bibitem{2} K. Binder and A. P. Young, Spin glasses: Experimental facts, theoretical concepts, and open questions, Rev. Mod. Phys. 58, 801 (1986).
\bibitem{3} C. A. Angell, Formation of glasses from liquids and biopolymers, Science 267, 1924 (1995).
\bibitem{4} L. Berthier and G. Biroli, Theoretical perspective on the glass transition and amorphous materials, Rev. Mod. Phys. 83, 587 (2011).
\bibitem{5} K. Amann-Winkel, R. Böhmer, F. Fujara, C. Gainaru, B. Geil, and T. Loerting, Colloquium: Water's controversial glass transitions, Rev. Mod. Phys. 88, 011002 (2016).
\bibitem{6} K. H. Fischer and J. A. Hertz, Spin Glasses (Cambridge University Press, Cambridge, 1993).
\bibitem{7} M. Goldstein, Viscous liquids and the glass transition: A potential energy barrier picture, J. Chem. Phys. 51, 3728 (1969).
\bibitem{8} L. C. E. Struik, Physical Aging in Amorphous Polymers and Other Materials (North-Holland, Amsterdam, 1978).
\bibitem{9} G. Adam and J. H. Gibbs, On the temperature dependence of cooperative relaxation properties in glass-forming liquids, J. Chem. Phys. 43, 139 (1965).
\bibitem{10} R. A. L. Jones, Soft Condensed Matter, Oxford Master Series in Condensed Matter Physics Vol. 6 (Oxford University Press, Oxford, 2002).
\bibitem{11} W. Gotze and L. Sjogren, Relaxation processes in supercooled liquids, Rep. Prog. Phys. 55, 241 (1992).
\bibitem{12} N. H. March and M. P. Tosi, Introduction to Liquid State Physics (World Scientific, Singapore, 2002).
\bibitem{13} W. Kauzmann, The nature of the glassy state and the behavior of liquids at low temperatures, Chem. Rev. 43, 219 (1948).
\bibitem{14} G. S. Fulcher, Analysis of recent measurements of the viscosity of glasses, J. Am. Ceram. Soc. 8, 339 (1925).
\bibitem{15} P. G. Debenedetti, F. H. Stillinger, and M. S. Shell, Model energy landscapes, J. Phys. Chem. B 107, 14434 (2003).
\bibitem{16} M. S. Shell, P. G. Debenedetti, E. La Nave, and F. Sciortino, Energy landscapes, ideal glasses, and their equation of state, J. Chem. Phys. 118, 8821 (2003).
\bibitem{17} C. Scalliet, L. Berthier, and F. Zamponi, Nature of excitations and defects in structural glasses, Nat. Commun. 10, 5102 (2019).
\bibitem{18} Y. He and V. Lubchenko, Knowledge as a breaking of ergodicity, Neural Comput. 37, 742 (2025).
\bibitem{19} F. Schwabl, Statistical Mechanics (Springer, Berlin, 2002).
\bibitem{20} M. Kardar, Statistical Physics of Particles (Cambridge University Press, Cambridge, 2007).
\bibitem{21} S. Akhanjee, Distribution function for $n \ge g$ quantum particles, arXiv:2411.09877.
\bibitem{22} R. P. Stanley, Enumerative Combinatorics, 2nd ed. (Cambridge University Press, Cambridge, 2012), Vol. 1.
\bibitem{23} D. Zwillinger, CRC Standard Mathematical Tables and Formulae, 30th ed. (CRC Press, Boca Raton, FL, 1996).
\bibitem{24} C. P. Winsor, The Gompertz curve as a growth curve, Proc. Natl. Acad. Sci. USA 18, 1 (1932).
\bibitem{25} A. Hansen, E. G. Flekky, and B. Baldelli, Anomalous diffusion in systems with concentration-dependent diffusivity: Exact solutions and particle simulations, Front. Phys. 8, 519624 (2020).
\bibitem{26} G. H. Hardy and S. Ramanujan, Asymptotic formul\ae{} in combinatory analysis, Proc. London Math. Soc. s2-17, 75 (1918).
\bibitem{27} E. W. Weisstein, CRC Concise Encyclopedia of Mathematics (CRC Press, Boca Raton, FL, 2002).
\bibitem{28} J. D. Bekenstein, Black holes and entropy, Phys. Rev. D 7, 2333 (1973).
\bibitem{29} S. W. Hawking, Particle creation by black holes, Commun. Math. Phys. 43, 199 (1975).
\bibitem{30} V. G. Knizhnik and A. B. Zamolodchikov, Current algebra and Wess-Zumino model in two dimensions, Nucl. Phys. B 247, 83 (1984).
\bibitem{31} P. Francesco, P. Mathieu, and D. Sénéchal, Conformal Field Theory, Graduate Texts in Contemporary Physics (Springer, New York, NY, 2012).
\bibitem{32} J. L. Cardy, Operator content of two-dimensional conformally invariant theories, Nucl. Phys. B 270, 186 (1986).
\bibitem{33} B. Derrida, Random-energy model: Limit of a family of disordered models, Phys. Rev. Lett. 45, 79 (1980).
\bibitem{34} K. Shing and K. Gubbins, The chemical potential in dense fluids and fluid mixtures via computer simulation, Mol. Phys. 46, 1109 (1982).
\bibitem{35} L. Kosztin, B. Barz, and L. Janosi, Calculating potentials of mean force and diffusion coefficients from nonequilibrium processes without Jarzynski's equality, J. Chem. Phys. 124, 064106 (2006).
\bibitem{36} W. Jin, P. Vashishta, R. K. Kalia, and J. P. Rino, Dynamic structure factor and vibrational properties of SiO$_2$ glass, Phys. Rev. B 48, 9359 (1993).
\bibitem{37} A. C. Hewson, The Kondo Problem to Heavy Fermions, (Cambridge University Press, Cambridge, 1997).
\bibitem{38} S. Ouvry and A. P. Polychronakos, Inclusion statistics, arXiv:2511.22710.
\bibitem{39} F. D. M. Haldane, ``Fractional statistics'' in arbitrary dimensions: A generalization of the Pauli principle, Phys. Rev. Lett. 67, 937 (1991).
\bibitem{40} F. Wilczek, Directions in Condensed Matter Physics (World Scientific, Singapore, 1990), Vol. 10.
\end{thebibliography}
\end{document}